\newcommand{\be}[0]{\begin{equation}}
\newcommand{\ee}[0]{\end{equation}}
\newcommand{\ba}[0]{\begin{eqnarray}}
\newcommand{\ea}[0]{\end{eqnarray}}

\documentclass[11pt,a4,axodraw]{iopart}
\usepackage{axodraw}
\usepackage{amssymb}
\usepackage{epsfig}

\begin{document}

\title{Low $Q^2$ proton structure function, using gluon and pseudoscalar meson clouds in the constituent quark framework }

\author{A.Mirjalili $^{*}$$^{,1}$$^{,4}$, M.M.Yazdanpanah $^{2}$$^{,4}$, F.Taghavi-Shahri $^{4}$and K.Ghorbani $^{3}$$^{,4}$}

\address{$^{1}$ Physics Department, Yazd University, 89195-741, Yazd, Iran\\$^{2}$ Physics
 Department, Kerman Shahid Bahonar University, Kerman, Iran\\ $^{3}$ Physics Department, Arak University
,Arak, Iran\\$^{4}$ School of Particles and Accelerators, IPM(Institute for research in fundamental sciences, 19395-5531, Tehran, Iran}
\ead{$^{*}$Mirjalili@ipm.ir}
\begin{abstract}
The idea of the meson cloud approach in the chiral quark model has
been extended to include gluon cloud in order  to achieve the
parton densities in the nucleon, based on the constitute quark
framework. The  splitting function of the  quark to the
quark-meson and quark-gluon  at low $Q^2$ value are used to obtain
parton densities in the constituent quark. The phenomenological
constituent model is employed to extract the parton distributions
in the proton at low $Q^2$ value. Since we have access to the
parton densities at low $Q^2$, we are able to obtain
$F_{2}(x,Q^2)$ structure function at low $Q^2$ value. The result
is in good agreement with available experimental data and some
theoretical models.  To confirm the validity of our calculations,
the fraction of total momentum of proton which is carried
by gluon at high $Q^2$ and also the Gottfried sum rule are computed.
The results are in good agreement
with what are expected.
\end{abstract}
\section{Introduction}
In hard scattering events, mesons and baryons can be viewed as
bound states built up from partonic constituents, i.e. quarks and
gluons. This picture changes at low energies, where hadronic
effects play a more prominent rˆole in the non-perturbative
structure of hadrons. One particular importance is the pion cloud
effects which e.g. have a direct impact on the spin structure of
the proton \cite{Thomas}. Thus they need to be incorporated in
bound-state calculations aiming at a realistic description of
mesons and baryons. Pion effects on the quark propagation are
important for several reasons. They account for (at least part of
the) pion cloud effects in baryons and mesons. Furthermore they
allow for the possibility of hadronic intermediate states in bound
state calculations and therefore
generate the finite width of meson spectral functions. \\

Dynamical chiral symmetry breaking is one of the most important
properties of low energy QCD. The breaking pattern has profound
impact for phenomenological quantities, as e.g. the appearance of
the pseudoscalar Goldstone bosons in the chiral limit of QCD and
the non-degeneracy of chiral partners. Chiral perturbation theory
~\cite{Weinberg,Gasser} describes these effects very efficiently
on the level of hadrons but has nothing to say about the
underlying structure of the full theory.  The interplay between
the fundamental quark and gluon degrees of freedom and the
resulting bound states are also particularly interesting. In full
QCD there are hadronic contributions to the fully dressed
quark-gluon interaction. These effects are generated by the
inclusion of dynamical sea quarks in the quark-gluon interaction
and are therefore only present in unquenched QCD. The quark-gluon
vertex is also an important ingredient into the quark-antiquark
interaction that is responsible for the formation and properties
of bound states. On a perturbative level, the quark-gluon vertex
has been studied in detail in arbitrary gauge and dimensions in
\cite{Davydychev}. However
nonperturbative properties of this vertex are also still under tense scrutiny. \\

At low energies, the idea that baryons are made up of three
constituent quarks and mesons of a (constituent) quark-antiquark
pair , the naive quark model scenario, accounts for a large number
of experimental facts . The quest for a relation between the two
regimes, i.e. between the current quarks of the theory and the
constituent quarks of the model has an old history  and, in recent
years, this search has been the subject of a considerable research
effort . The fundamental problem one would like to understand is
how confinement, i.e. the apparent absence of color charges and
dynamics in hadron physics, is realized. Detailed quark models of
hadron structure based on the constituent quark concept have been
defined in order to explain low energy properties \cite{Isgur}. To
proceed from these models to the asymptotic regime, where deep
inelastic scattering (DIS) takes place, a hadronic scale is
associated to the model calculations. The experimental conditions
are reached by projecting the leading twist component of the
observable and evolving according to perturbative QCD. The
procedure describes successfully the gross features of the DIS
results. It was long ago, at the time that QCD was being proposed,
that a procedure, hereafter called ACMP(Altarelli, Cabibo, Miani,
Petronzio) \cite{ACMP}, was developed to understand the relation
between the constituent quarks and the partons. In this approach,
constituent quarks are complex objects, made up of point-like
partons (current quarks, antiquarks and gluons), interacting by a
residual interaction described by a quark model. The hadron
structure functions are obtained as a convolution of the
constituent quark wave function with the constituent quark
structure function. This procedure has been recently reviewed to
estimate the structure function of the pion with success. In the
ACMP approach, each constituent quark is dressed by a neutral
cloud of quark-antiquark pairs and gluons, thus, this scenario
supports the confinement mechanism. A few years earlier a second
approach had been developed \cite{Kuti}, in which the proton is
assumed to be made out of three valence quarks plus a neutral core
of quark-antiquark pairs and gluons, very much in the spirit of
recent developments along the Manohar-Georgi model \cite{Manohar}.
This duality of approaches has to do, in modern language, with the
implementation of Chiral Symmetry Breaking(CSB). The naive models
do not contain spontaneous CSB and this phenomenon has to be
implemented if
they are to represent QCD at low energies. \\

The effective chiral quark theory \cite{Manohar} may provide an
alternative explanation to that of the traditional meson cloud
approach \cite{main9}. In this theory, the relevant degrees of
freedom are constituent quarks, gluons and Goldstone bosons. The
chiral quark model ($\chi QM$) includes both gluon and pion
exchange between constituent quarks together with corresponding
exchange currents. The relevant degrees of freedom and the related
question whether the pions couple effectively to the nucleon or to
the constituent quarks is extensively
discussed in Refs.\cite{main9,main14}. It is necessary to study the consequences of these different
scenarios in a broad range of
physical processes to assess their validity.\\

On the other hand, the $\chi QM$ can be used to study the flavor
structure of the constituent quark model and the nucleon within
the conventional mesonic cloud picture. Using this model the
effects of $SU(3)_f$ symmetry breaking can be discussed
\cite{main}. The implications of the Gottfried sum rule (GSR)
violation for the $\Delta$-n mass splitting were also considered
in Ref.\cite{main}. At  low energy resolution scale the
constituent quark picture successfully describes hadronic
structure functions. The sea quark and gluonic degrees of freedom
are assumed to be absorbed into constituent quarks to be
considered as quasi-particles \cite{ymain2}. A relation between
the two regimes of hadron structure function description; i.e. the
chiral quark and the constituent quark models, has a considerable
significance which has been investigated widely in the literature,
and has attracted much attention in recent years \cite{ymain3}.\\

As should be noticed, the main ingredients of this paper are two
subjects.  In continuation of our previous work \cite{IJMPA08} we
add the gluon cloud to the  $\chi$QM while we use an effective
lagrangian at low $Q^2$ values. We resort to a constituent quark
model to extract parton densities inside the proton. Since the
gluon densities are also at our access, we are able to calculate
$F_2$ structure function
for the proton at the NLO approximation.\\

The organization of the paper is as follows: In section 2 we
introduce Quark-meson and in similar way quark-gluon vertex
function based on nucleonic Sullivan deep-inelastic scattering
\cite{Sullivan}. Section 3 is allocated to $\chi QM$ and the
constituent quark distribution  is obtained. In addition we
consider the  gluon cloud in the constituent quark. Therefore  we
can calculate the gluon distribution function inside the proton.
This contribution has not been considered in Ref.\cite{main}. In
this section  we also discuss about phenomenological valon model
\cite{Hwa 2002} to extract valence distribution
in the meson. These distributions are required to obtain mesonic
anti-quark contribution in the constituent quarks. In section 4 we
present our result for $F_2$ structure function at the NLO
approximation, using the  parton densities in the proton which are
extracted from $\chi QM $. To confirm the validity of  our
calculation, we evolve the gluon distribution to high $Q^2$ to get
the momentum fraction of proton which is carried by gluon. Using
the antisymmetric  property of sea quark densities
which is resulted from  $\chi QM $, we also
calculate the Gottfired sum rule to test again the validity of our
calculations. The conclusion is given in section
5.
\section{Chiral quark model and effective quark-meson and  quark-gluon interactions}

This model was introduced by Georgi and Manohar~\cite{Manohar} in
order to incorporate the chiral symmetry of QCD into the
successful features of the constituent quark model~\cite{DGG}. The
prime assumption of the model is the possible realization of an
effective Lagrangian between the scale of chiral symmetry breaking
$\Lambda_{\chi}$ and confinement scale $\Lambda_{QCD}$. The
dynamical degrees of freedom here are constituent quarks,
pseudoscalar mesons and arguably gluons. The respecting Lagrangian
can be written as \ba \label{lag1} \mathcal{L}_{eff} = i{\bar Q}
\gamma_{\mu}(\partial^{\mu}+ig_{s}G^{\mu})Q
-\frac{g_{A}}{f_{\pi}} {\bar Q} \partial_{\mu} U \gamma^{\mu} \gamma_{5} Q \nonumber\\
-M_{Q} {\bar Q}Q +\frac{f_{\pi}}{4} trac( D_{\mu}U
D^{\mu}U^{\dagger})-\frac{1}{4} {G^{a}_{\mu\nu}} { G^{\mu\nu}_{a}}
\,. \ea The matrix $U$ contains the pseudoscalar mesons and $Q$
stands for the constituent quark. $G^{\mu\nu}$ is the strength
field tensor defined as following
\begin{equation}
G^{\mu\nu,a} = \partial^{\mu}G^{\nu,a} -\partial^{\nu} G^{\mu,a} -
if^{abc} G^{\mu}_b G^{\nu}_c\,
\end{equation}
where $G^{\mu}$ is the gluon field and f's are the structure
constants. The covariant derivative, $D^{\mu}$, is associated to
the chiral symmetry of QCD in flavour space.

$M_{Q}$, $f_{\pi}$ and $g_{A}$ are the constituent quark mass, the
pion decay constant and the axial-vector constant, respectively.
The strong running coupling, $g_{s}$, has to be considered in some
energy below $\Lambda_{\chi}$
and we then take it as a constant.\\
\subsection{Quark-meson effective vertex function}

The effective chiral quark model \cite{Manohar} is applied in
order to study the pseudoscalar meson clouds in the constituent
quarks \cite{main}. In Ref.\cite{main} it is found that the pionic
clouds can explain the violation of the Gottfried sum rule and
introduce an enhancement on the non-perturbative effects of the
sea quark pairs. To this end the pion-quark splitting function is
introduced in analogy to the nucleonic Sullivan deep-inelastic
scattering and expressed in \cite{Sullivan} as: \ba
f_{Q\rightarrow MQ'}(x_M,k_{\bot}^2)&=&{g_{Q\rightarrow MQ'}^2
\over 16\pi^2}{1\over x_M(1-x_M)}\left| G_{Q\rightarrow
MQ'}(x_M,k_{\bot}^2) \right|^2\nonumber\\
&&\times{((1-x_M)m_Q-m_{Q'})^2+k_{\bot}^2 \over
(1-x_M)(m_Q^2-M_{MQ'}^2)^2},\label{split} \ea where $x_M$ is the
longitudinal (light cone) momentum fraction  of the constituent
quark for the Goldstone boson and $k_{\bot}$ is the perpendicular
momentum of the quark $Q'$. The $g_{_{Q\rightarrow MQ'}}$ is the
effective coupling constant of pion-constituent quark:
\begin{equation}
g_{_{Q\rightarrow MQ'}}^2 = {g_A^2\over f^2}{(m_Q+m_{Q'})^2\over
4},
\end{equation}
where  $g_A$ is the axial vector coupling constant of the
constituent quark which is equal to one. We take
$m_l=m_{Q'}=\frac{m_N}{3}=313$ $MeV $ for the light up and down
quarks and $m_s=m_{Q'}=m_{\Sigma}-m_{N}+m_{l}=567$ $MeV $
for the strange quarks.\\
$M_{MQ'}^2$ is the invariant mass squared of $MQ'$ system which
 is defined as:
\begin{equation}
M_{MQ'}^2 = {m_M^2+k_{\bot}^2 \over x_M}+{m_{Q'}+k_{\bot}^2 \over
1-x_M}.
\end{equation}
 The $G_{Q\rightarrow MQ'}$ is a vertex function or
phenomenological form factor for which we adopt the exponential
form:
\begin{equation}
G_{Q\rightarrow MQ'} = \exp \left ( {m_Q^2-M_{MQ'}^2
(x_M,k_{\bot}^2)\over 2\Lambda_{\chi}^2} \right ).\label{vertex}
\end{equation}
$\Lambda_{\chi}$ is a cut-off parameter which can be taken equal
for all fluctuations involving pseudoscalar or vector mesons. The
integration of the splitting function over $k_{\bot}$
$[f_{Q\rightarrow MQ'}(x_{_M})= \int_0^{\infty}f_{Q\rightarrow
MQ'}(x_{_M}, k^2_T)\;dk^2_T ]$ and then over $x_M$ and finally summing
over the intermediate quarks ($Q'$) yields:
\begin{equation}
P_{M/Q}=|a_{M/Q}|^2=\sum_{Q'} \int_0^1 f_{Q\rightarrow
MQ'}(x_{_M})dx_{_M},\label{Integrad-split}
\end{equation}
which is the probability of finding a Goldstone boson $M$ in the
constituent quark $Q$.\\

\subsection{Quark-gluon effective vertex function}
Fairly  gluon distributions can be obtained by dressing quarks
with gluons in the nonperturbative regime with massive
 effective gluons  ($m^{eff}_g$ ) and frozen running $\alpha_s$.
Rather heavy effective gluons $m^{eff}_g$ $>$ 0.4 $GeV$ and small
$\alpha_s$ $<$ 0.5 are required in order to limit the momentum
carried by quarks to approximately what is  required by the
phenomenology \cite{GRV}. Now in order to include the gluon clouds
in the constituent quarks we need to put the line of analogy some
further, in the sense that an almost the same form of splitting
function is regarded for the gluon-quark interaction as that for
the quark-meson interaction. The main differences stand on two
parts. The first one is the quark-meson coupling constant, that we
replace it with the strong coupling constant at some low energy.
Secondly, we need to know the relevant vertex function for the
quark-gluon interaction. The vertex function encodes the extended
structure of the gluon and the constituent quarks. The extraction
of the vertex function is rather difficult since it incorporates
the non-perturbative effects. However, in a series of recent
studies \cite{Fischer2009,Fischer2008}, the authors have
calculated the non-perturbative corrections to the quark-gluon
vertex in the framework of the Dyson-Schwinger and Bethe-Salpeter
equation. Their predictions for the light meson properties seems
satisfactory\cite{Fischer2009}. On the other hand, we have found
out that our ansatz for the quark-gluon vertex  which is assumed
similar to the quark-meson vertex,  has qualitatively the same
momentum behavior.

Consequently, we have the quark-gluon fluctuations which tends to
the  following  splitting function: \ba
f_{Q\rightarrow gQ'}(x_g,k_{\bot}^2)&=& {\alpha_s(Q^2) \over
4\pi}{1\over x_g(1-x_g)}\left| G_{Q\rightarrow
gQ'}(x_g,k_{\bot}^2) \right|^2\nonumber\\
&&\times{((1-x_g)m_Q-m_{Q'})^2+k_{\bot}^2 \over
(1-x_g)(m_Q^2-M_{gQ'}^2)^2},\label{split} \ea where $x_g$ is the
longitudinal (light cone) momentum fraction  of the constituent
quark for the gluon  and $k_{\bot}$ is the perpendicular momentum
of the quark $Q'$.\\

The integration of the quark-gluon  splitting function over
$k_{\bot}$  and then over $x_g$ and finally summing over the
intermediate quarks ($Q'$) yields:
\begin{equation}
P_{g/Q}=|a_{g/Q}|^2=\sum_{Q'} \int_0^1 f_{Q\rightarrow
gQ'}(x_{_g})dx_{_g},\label{Integrad-split}
\end{equation}
\section{Constituent quark distribution function in the chiral quark model}
The constituent quark Fock-state $|Q\rangle$, can be expressed in
terms of a series of light-cone Fock-states:
\begin{equation}
|Q\rangle = \sqrt{Z}|q\rangle + \sum_{q'}a_{{\cal B}/Q}|q', {\cal
B}\rangle\label{Q},
\end{equation}
where $|q\rangle$ is the ``bare'' but massive state, $\sqrt{Z}$
denotes the renormalization factor for a ``bare'' constituent
quark and $|a_{{\cal B}/Q}|^2$ are probabilities to find Goldstone
bosons and gluon distribution  in the constituent quark states.
Then the dressed u- and d-quark Fock-states are:
\begin{eqnarray}
|U\rangle & = & \sqrt{Z}|u\rangle +\sqrt{1\over 3}\ a_{\pi^0/U}|u,
\pi^0\rangle +\sqrt{2\over 3}\ a_{\pi^+/U}|d,\pi^+\rangle
+a_{k^+/U}|s, K^+\rangle\nonumber\\ &&+\ a_{g/U}|u,
g\rangle+\cdots,\\
|D\rangle & = & \sqrt{Z}|d,\rangle +\sqrt{1\over 3}\
a_{\pi^0/D}|d,\pi^0\rangle +\sqrt{2\over 3}\
a_{\pi^-/D}|u,\pi^-\rangle +a_{k^0/D}|s,K^0\rangle \nonumber\\
&&+\ a_{g/D}|d, g\rangle+\cdots.
\end{eqnarray}
The above expressions can be depicted as in the following graphs:

\begin{center}
\begin{picture}(400,50)(-55,0)
\label{graph} \SetColor{Black} \SetScale{1}
 {\SetWidth{1.5}\Line(0,10)(30,10)
\Text(15,5)[t]{\small{$U$}}} \PText(35,18)(0)[t]{=}
   \Line(40,10)(70,10)
\Text(55,5)[t]{\small{$u$}} \PText(75,18)(0)[t]{+}
\Text(83,5)[t]{\small{$u$}} \Text(100,5)[t]{\small{$u$}}
\Text(117,5)[t]{\small{$u$}}
    \Line(80,10)(120,10)
\PText(125,18)(0)[t]{+}
    \DashCArc(100,10)(12,0,180){2}
\Text(105,35)[t]{\small{$\pi^{0}$}}
    \Vertex(88,10){1.5}
    \Vertex(112,10){1.5}
    \Line(130,10)(170,10)
    \DashCArc(150,10)(12,0,180){2}
    \Vertex(138,10){1.5}
    \Vertex(162,10){1.5}
\Text(155,35)[t]{\small{$\pi^{+}$}} \PText(175,18)(0)[t]{+}
\Text(133,5)[t]{\small{$u$}} \Text(151,6)[t]{\small{$d$}}
\Text(168,5)[t]{\small{$u$}}
    \Line(180,10)(220,10)
    \DashCArc(200,10)(12,0,180){2}
    \Vertex(188,10){1.5}
    \Vertex(212,10){1.5}
   \Line(180,10)(220,10)
    \DashCArc(200,10)(12,0,180){2}
    \Vertex(188,10){1.5}
    \Vertex(212,10){1.5}
\Text(205,35)[t]{\small{$K^{+}$}} \PText(225,18)(0)[t]{+}
    \Line(230,10)(270,10)
\Text(183,5)[t]{\small{$u$}} \Text(201,6)[t]{\small{$s$}}
\Text(218,5)[t]{\small{$u$}}
    \GlueArc(250,10)(12,0,180){2}{8}
\Text(250,35)[t]{\small{$g$}}
   \Vertex(238,10){1.5}
    \Vertex(262,10){1.5}
\Text(233,5)[t]{\small{$u$}} \Text(250,6)[t]{\small{$u$}}
\Text(268,5)[t]{\small{$u$}} \PText(280,18)(0)[t]{+ ...}
\SetColor{Black}
\SetColor{Black} \SetScale{1}{\SetWidth{1.5}
 \Line(0,-45)(30,-45)
\Text(15,-50)[t]{\small{$D$}}} \PText(35,-37)(0)[t]{=}
   \Line(40,-45)(70,-45)
\Text(55,-50)[t]{\small{$d$}} \PText(75,-37)(0)[t]{+}
\Text(83,-50)[t]{\small{$d$}} \Text(100,-50)[t]{\small{$d$}}
\Text(117,-50)[t]{\small{$d$}}
    \Line(80,-45)(120,-45)
\PText(125,-37)(0)[t]{+}
    \DashCArc(100,-45)(12,0,180){2}
\Text(105,-20)[t]{\small{$\pi^{0}$}}
    \Vertex(88,-45){1.5}
    \Vertex(112,-45){1.5}
    \Line(130,-45)(170,-45)
    \DashCArc(150,-45)(12,0,180){2}
    \Vertex(138,-45){1.5}
    \Vertex(162,-45){1.5}
\Text(155,-20)[t]{\small{$\pi^{+}$}} \PText(175,-37)(0)[t]{+}
\Text(133,-50)[t]{\small{$d$}} \Text(151,-49)[t]{\small{$u$}}
\Text(168,-50)[t]{\small{$d$}}
   \Line(180,-45)(220,-45)
    \DashCArc(200,-45)(12,0,180){2}
    \Vertex(188,-45){1.5}
    \Vertex(212,-45){1.5}
\Text(205,-20)[t]{\small{$K^{0}$}} \PText(225,-37)(0)[t]{+}
    \Line(230,-45)(270,-45)
\Text(183,-50)[t]{\small{$d$}} \Text(201,-49)[t]{\small{$s$}}
\Text(218,-50)[t]{\small{$d$}}
    \GlueArc(250,-45)(12,0,180){2}{8}
\Text(250,-20)[t]{\small{$g$}}
   \Vertex(238,-45){1.5}
    \Vertex(262,-45){1.5}
\Text(233,-50)[t]{\small{$d$}} \Text(250,-49)[t]{\small{$d$}}
\Text(268,-50)[t]{\small{$d$}} \PText(280,-37)(0)[t]{+ ...}
\SetColor{Black}
 \end{picture}
\end{center}
\vspace{2cm}

In these graphs  the thick lines indicate the quark propagators,
dashed lines are meson fields and
wiggly curves stand for the gluons, respectively.\\[.5cm]

We consider the nucleon to be a bound state of three constituent
quarks ($U$ and $D$). The quark distributions in the constituent
quark, at some QCD initial scale, can be written as:
\begin{eqnarray}
u_U(x)&=& u_U^{(0)}(x)+u_U^{(i)}(x)+u_U^{(\pi)}(x)+u_U^{(g)}(x),\label{uu}\\
 u_D(x)&=& u_D^{(i)}(x)+u_D^{(\pi)}(x),\label{ud}\\
d_D(x)&=& d_D^{(0)}(x)+d_D^{(i)}(x)+d_D^{(\pi)}(x)+d_D^{(g)}(x),\label{dd}\\
 d_U(x)&=& d_U^{(i)}(x)+d_U^{(\pi)}(x),\label{du}\\
s_{U(D)}(x)&=& s_{U(D)}^{(i)}+s_{U(D)}^{K}(x),\label{sd}\\
g_{U}(x)&=&u_U^{(g)}(1-x)\label{gu}\\
g_{D}(x)&=&d_D^{(g)}(1-x)\label{gd}\;.
\end{eqnarray}
The anti-quark distributions become:
\begin{eqnarray}
\bar{u}_U(x)= & \bar{u}_U^{(\pi)}=\bar{d}_D(x) & =\bar{d}_D^{(\pi)},\\
\bar{u}_D(x)= & \bar{u}_D^{(\pi)}=\bar{d}_U(x) & =\bar{d}_U^{(\pi)},\\
\bar{s}_U(x)= & \bar{s}_U^{(K)}=\bar{s}_D(x) & =\bar{s}_D^{(K)},\\
\tilde{u}_U(x)= & \tilde{u}_U^{(k)}=\tilde{d}_D^{(k)}\;.
\end{eqnarray}
It should be noted that for sea quark densities we have
${q}_Q^{(M)}=\bar{q}_Q^{(M)}$. The superscripts denoted with $(0)$
correspond to the ``bare'' quark distributions, and those denoted
with ($i$ and $g$) to the intermediate quark distributions
associated with mesons and gluons respectively, and those denoted
with $(\pi)$ originate from mesons (pions).

The bare quark distribution in the constituent quarks has the
form:
\begin{equation}
u_U^{(0)}(x)=d_D^{(0)}(x)=\left ( 1-\sum_{\cal C} P_{{\cal C}/Q}
\right )\delta (x-1)\;, \label{bare-valence}
\end{equation}
where these distributions play the role of the valance quark
distributions inside the constituent quarks. In Eq.~(\ref{bare-valence}), $P_{\cal C}/Q$ refers  to the probability of finding a Goldstone boson and gluon in the constituent quark Q. So back to Eq.(\ref{Q}), we have $P_{\cal C}/Q=|a_{{\cal B}/Q}|^2$.\\

The intermediate quark distribution function in the constituent
quark is calculated from the meson splitting function:
\begin{eqnarray}
u_U^{(i)}(x)= & d_D^{(i)}(x) & ={1\over 3}f_{\pi/Q}(1-x),\label{intermediate1}\\
u_D^{(i)}(x)= & d_U^{(i)}(x) & ={2\over 3}f_{\pi/Q}(1-x),\label{intermediate2}\\
s_U^{(i)}(x)= & s_D^{(i)}(x) &
=f_{K/Q}(1-x)\label{intermediate3}\;,
\end{eqnarray}
where $f_{\frac{M}{Q}}$ is the total splitting function of the
constituent quark and is defined as \ba f_{\frac{M}{Q}}\equiv
\sum_{Q^{'}}f_{Q\rightarrow{MQ^{'}}}\;. \ea
 Mesonic anti-quark contributions
in the constituent quarks which finally yield us sea quark
distributions inside the constituent quarks, are given by the
equations:
\begin{equation}
\bar{u}_U^{(\pi)}(x)={1\over 6}I_{\pi}(x), \quad
\bar{u}_D^{(\pi)}(x)={5\over 6}I_{\pi}(x), \quad
\bar{s}_U^{(K)}(x)=I_K(x),\label{mIm}
\end{equation}
where
\begin{equation}
I_M(x)=\int_x^1 f_{M/Q}(y)q_M \left ( {x\over y} \right ){dy \over
y}\;.\label{Im}
\end{equation}

Here $q_M ( {x\over y})$ denotes the valence quark distribution of
the meson. These valence distributions  are required to extract
sea quark densities in the constituent quarks of the proton. They will be obtained in the
next section, using the phenomenological valon model.\\

 Using the constituent quark
model for the proton , the  parton  densities  at low $Q^2$ value,
for instance 0.5 $GeV^2$ can be obtained.
 More details regarding  the employed constituent model shall be explained in
 section.4.
\subsection{Valon model}
According to the valon model \cite{Hwa 2002}, a valon is a dressed
valence quark so that there is a one-to-one identification of a
valon with the associated valence quark as probed at high $Q^2$ .
In this model a meson, for instance,  is a bound state of two
valons. They contribute independently in an inclusive hard
collision with a $Q^2$-dependence that can be calculated in QCD at
high $Q^2$. The valon picture suggests that the structure function
of a meson involves a convolution of two distributions: the valon
distribution in the meson and the structure function for each
valon, so that one has
\begin{equation}
F_2^M(x,Q^2)=\sum_v \int_x^1 dy \, G_{v/M}(y) \, F_2^v(z={x\over
y},Q^2)\, , \label{conv-int}
\end{equation}
where the summation is over the two valons. Here $F_2^M(z,Q^2)$ is
the meson structure function, $F_2^v$ is the corresponding
structure function of a $v$ valon, and $G_{v/M}(y)$ indicates the
probability for the $v$ valon to have momentum fraction $y$ in the
meson. We shall assume that the two valons carry all the momentum
of the meson.\\
We assume the following  simple form for the exclusive valon
distribution inside the mesons which facilitates the
phenomenological analysis,
\begin{equation}
G_{v}(y_1, y_2)=g \, (y_1)^{p} \, y_2^{q} \, \delta (y_1+y_2-1),
\end{equation}
where $p$ and $q$ are two free parameters and $y_i$ is the
momentum fraction of the i'th valon. The $U$ and $D$ type
inclusive valon distributions can be obtained by  integration over
the specified variable,
\begin{equation}
G_{v_1}(y)=\int dy_2\,G_{v}(y, y_2)=g\, y^{p}\, (1-y)^{q},
\end{equation}
\begin{equation}
G_{v_2}(y)=\int dy_1\,G_{v}(y_1, y)=g\, y^{q}\, (1-y)^{p}.
\end{equation}
The normalization parameter $g$ has been fixed by requiring
\begin{equation}
\int_0^1 G_{v_1}(y) \, dy=\int_0^1 G_{v_2}(y) \, dy=1,
\end{equation}
and is given by $g=\frac{1}{B(p +1, q +1)}$, where $B(m, n)$  is the Euler-beta function.\\
Consequently, we will get the following inclusive valon
distributions for mesons:
 \ba
 G_{v_1}(y)&=&\frac{1}{B(p+1,q+1)}y^p (1-y)^q\nonumber\;,\\
 G_{v_2}(y)&=&\frac{1}{B(q+1,p+1)}y^q (1-y)^p\;.\label{valon}
 \ea
The  dirac delta function, $\delta(y_1+y_2-1)$, automatically
ensures the momentum sum rule:
 \ba
\int_0^1 y G_{v_1} dy+\int_0^1 y G_{v_2} dy=1\;, \ea Mellin
transformation from Eq.~(\ref{valon}), will yield the
 moment distributions for valons \cite{IJMPA08}. The
 moments of quark and gluon distributions at any energy scale inside the meson are obtained by multiplying
 the valon moments with the appropriate moments of singlet, non-singlet and gluon sectors.
 Using the inverse Mellin transformation in the parameterized form as described in \cite{IJMPA07}
 and fitting over the available experimental data, the valence quark densities inside the mesons will be obtained.
\section{Results and discussions}
Using the $\chi QM$ we are able to extract the valence, sea and
gluon densities inside the constituent quarks. To access to parton
densities inside the proton, we employ a constituent model. We
need the quark distribution in a proton, $q_N(x)$, which can be
obtained using the convolution of the corresponding quark
distributions in the constituent quark ($q_{U,D}(x/y)$) with the
light-cone momentum distribution of the constituent quark in the
proton ($U(y),D(y)$), so that :
\begin{equation}
q_N(x)=\int_x^1 \left [ 2U(y)q_U \left ( {x\over y} \right
)+D(y)q_D \left ( {x\over y} \right ) \right ]{dy\over
y}.\label{distribute}
\end{equation}
 Eq.~(\ref{distribute}) is the basis for the constituent quark
model in which we can use to obtain the quark densities in a
proton. In our calculations, U(y) and D(y) are parameterized as:
\ba & U(y)&={ \beta (\alpha_Q+1,\beta_Q+1)}
y^{\alpha_Q}(1-y)^{B_Q}\nonumber\\
& D(y)&={\beta(\gamma_Q+1,\eta_Q+1)}
y^{\gamma_Q}(1-y)^{\eta_Q}\label{constituent} \ea where
$B(\alpha_Q+1,\beta_Q+1)$ and $B(\gamma_Q+1,\eta_Q+1)$ are the
Euler beta functions. The normalization coefficient
$\frac{1}{B(\alpha_Q+1,B_Q+1)}$ and
$\frac{1}{B(\gamma_Q+1,\eta_Q+1)}$ and finally the unknown
parameters which exist in Eq.~({\ref{constituent}}) have been
fixed by requiring the number sum rule for valance quark densities
inside the proton and also the momentum sum rule for the parton
densities
inside the proton. \\
\begin{figure}[htp]
\begin{center}
\includegraphics[width=10.5cm]{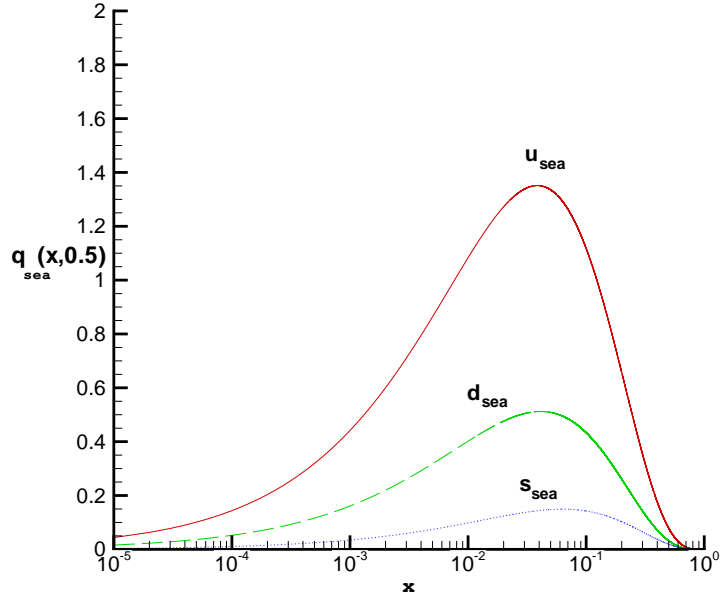}
\caption{Sea quark distribution in the proton at low $Q^2$= 0.5
$GeV^2$, based on $\chi QM$. The anti-symmetric of the sea quark
distributions are obvious.\label{fig1} }
\end{center}
\end{figure}

Requiring sum rules, we will get the following numerical values
for the parameters associated with the constituent quark
distributions :
$$
\begin{tabular}{|c|c|c|c|c|}
\hline $U(y)$       & $\alpha_U$    & $\beta_U$
\\  \hline
& -0.40 & 0.57  \\
\hline
$D(y)$       & $\gamma_U$    & $\eta_U$  \\
\hline
& -0.39 & 0.55  \\
\hline
\end{tabular}
$$
The obtained densities are at low $Q^2$ =$ 0.5 \;\;GeV^2$ which is
in correspond to  the chosen value for $\Lambda_x$=1.26 $GeV$ in
our calculations according to the model A of Ref.\cite{main}. The
results for sea and gluon densities in the
 proton are depicted in Fig.1 and Fig.2 respectively. The
 asymmetry of sea quark densities are obvious as we expected  from
 $\chi QM$.
\begin{figure}[htp]
\begin{center}
\includegraphics[width=10.5cm]{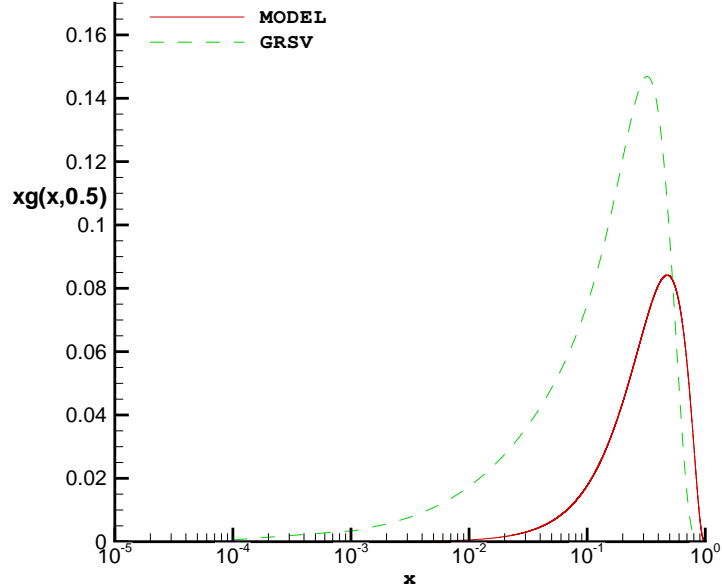}
\caption{Gluon distribution in the proton at low $Q^2$= 0.5
$GeV^2$, based on $\chi QM$. A comparison with GRSV has also been
done.\label{fig2} }
\end{center}
\end{figure}
By using the parton densities, the $F_2$ structure function at the NLO
approximation which
is defined by \ba\hspace{-1.7 cm} F_2^{ep}(x,Q^2)=x\sum_q
e_q^2\{q(x,Q^2)+{\bar
q}(x,Q^2)+\frac{\alpha_s(Q^2)}{2\pi}\times[C_{q,2}^{*}(q+\bar q)+
2\; C_{g,2}^{*} \;g]\label{GRV}\} \ea can be calculated. In Eq.(\ref{GRV}) $q,\bar q$ and $g$ refer
respectively to quark and gluon distribution inside the proton and
$C$ terms are Wilson coefficients which are defined in \cite{GRV}.
In Fig.3 the $F_2$ structure function for the proton  at low
$Q^2$= 0.5 $GeV^2$ is shown and compared with the GRSV model
\cite{GRV}. A comparison with available experimental data
\cite{data} has also been done. The agreement is well. Since we
have access to gluon distribution, to confirm the validity of
calculation at low $Q^2$ value, we can evolve it to high $Q^2$ and
calculate the fraction momentum of proton  carried by gluon. In
this regard we got 41.2{$\%$}  which is what we expect. Authors in
\cite{GRV} claimed that their extracted gluon distribution carries
about 50$\%$  of total momentum of the  proton. This is
more than what we got.
On viewing at Fig.2 which indicates gluon distribution in the proton at low $Q^2$, it
is predictable that we reach to lower presentation  of fraction momentum at evolved $Q^2$  value
in our model. A justifiable reason  for this  difference backs to this reality
that the results in \cite{GRV} is based on a global fit while we
employ $\chi QM$ to obtain the gluon contribution in our
calculation.\\

We use the following relation \cite{main} \ba\hspace{-1 cm}
S_G=\int_0^1
[F_2^{p}(x)-F_2^{n}(x)]\frac{dx}{x}=\frac{1}{3}+\frac{2}{3}\int_0^1(\bar{u}(x)-\bar{d}(x)){dx}=\frac{1}{3}-\frac{4}{9}P_\frac{\pi}{Q}\;,
\ea to obtain the Gottfried sum rule (GSR).
 The numerical value which
is obtained by this model is 0.2339 which is very near to quoted
experimental value $0.235\pm 0.026$ by NMC group
\cite{main2,main3}. Once again the validity of the calculation
using $\chi QM$
 at low $Q^2$
 value is confirmed.

\begin{figure}[htp]
\begin{center}
\includegraphics[width=10.5cm]{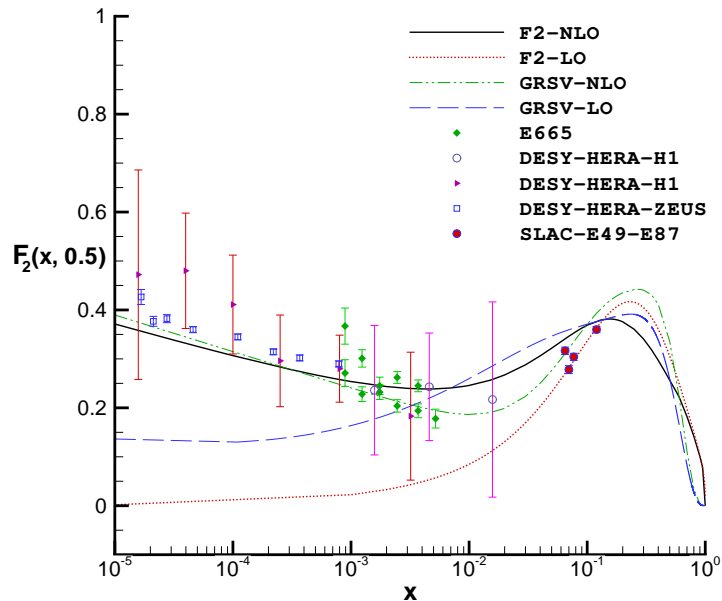}
\caption{\label{fig3}Analytical result for $F_2$ structure
function at LO and NLO approximation. The result have been
compared with GRSV model \cite{GRV} and available experimental
data \cite{data}.}
\end{center}
\end{figure}

\section{Conclusion}
The flavor structure of the nucleon in the effective chiral quark
model $(\chi QM)$ has been studied. In this model the Goldstone bosons couple
directly to the constituent quarks. This idea has been extended to
include gluon cloud in the  $\chi QM$  at low $Q^2$ value.
Consequently we could obtain an improvement
result for sea quark density and also calculated the gluon
distribution directly in $\chi QM$ while it has not been done in
previous works \cite{main,IJMPA08}. To obtain the sea quark
densities inside the constituent quark, we needed  valence quark
distributions of the meson . We got these valence distributions
using the phenomenological valon model \cite{Hwa 2002}. Furthermore the sea quark
densities  in the proton have been obtained, by convoluting the
required distributions in the used constituent quark model. This yielded a result in which the sea
quark densities in the proton are un-symmetrized. By accessing to the other partons in the proton,
we could calculate the $F_2$ structure
function at $Q^2$= 0.5 $GeV^2$ which confirms the anticipated result of
the model. For more validity of the model the fraction momentum of
proton which is carried by gluon at $Q^2$= 15 $GeV^2$ has been
calculated. The numerical result which was obtained for this
fraction at $Q^2$= 15 $GeV^2$ and also the numerical value for GSR
are very close to what are expected. However the obtained results
in this paper are satisfactory but one can use  different vertex
function, quoted in \cite{German people-Numerial assumption}. By
comparing the results, one  can choose  the best candidate to
consider meson and gluon clouds at low values  $Q^2$. This can be
done as a new research job in future. Further suggestion is to
consider valence quark density inside the constituent quark with a
gaussian form rather than the Dirac Delta as in
Eq.~(\ref{bare-valence}). In this case we would expect to achieve
 the number sum rules for the both constituent and valence quarks in a more
straightforward way. This will also be a scientific challenge for
a research task in future.
\section{Acknowledgment}
Authors are indebted to the Institute for Studies in Theoretical
Physics and Mathematics (IPM) for their hospitality whilst this
research was performed.

\end{document}